# Electrically Injected mid-infrared GeSn laser on Si operating at 140 K

Sudip Acharya, Hryhorii Stanchu, Rajesh Kumar, Solomon Ojo, Mourad Benamara, Guo-En Chang, Baohua Li, Wei Du, and Shui-Qing Yu,

*Abstract*— **Owing to its true direct bandgap and tunable bandgap energies, GeSn alloys are increasingly attractive as gain media for mid-IR lasers that can be monolithically integrated on Si. Demonstrations of optically pumped GeSn laser operating at room temperature under pulsed excitation and at cryogenic temperature under continuous-wave excitation show great promise of GeSn lasers to be efficient electrically injected light sources on Si. Here we report electrically injected GeSn lasers using Fabry-Perot cavity with 20 μm, 40 μm, and 80 μm ridge widths. A lasing threshold of 0.756 kA/cm² at 77 K, emitting wavelength of 2722 nm, and maximum operating temperature of 140 K were obtained. The lower threshold current density compared to previous works was achieved by reducing optical loss and improving the optical confinement. The peak power was measured as 2.2 mW/facet at 77 K.**

*Index Terms*— **Si photonics, GeSn laser, mid-infrared, electrically injected.**

## I. INTRODUCTION

SILICON photonics has recently been acknowledged as the key enabling technology for low-cost, chip-scale devices by leveraging the mature complementary metal-oxide-semiconductor (CMOS) processes [1–3]. Silicon photonics chips combine various building blocks including active and

Manuscript received ####; revised ####; accepted ####. Date of publication ####; date of current version ####. This work was supported by Air Force Office of Scientific Research (AFOSR) under grants FA9550-22-1-0493 and FA9550-19-1-0341. (Corresponding author: Shui-Qing Yu).

Sudip Acharya is with Material Science and Engineering, University of Arkansas, Fayetteville, AR 72701 USA and Department of Electrical Engineering and Computer Science, University of Arkansas, Fayetteville, AR 72701 USA (e-mail: sa066@uark.edu).

Hryhorii Stanchu, Rajesh Kumar, and Mourad Benamara are with the Institute for Nanoscience and Engineering, University of Arkansas, Fayetteville, AR 72701 USA (e-mails: hstanchu@uark.edu, mourad@uark.edu, rajeshk@uark.edu).

Solomon Ojo is with Material Science and Engineering, University of Arkansas, Fayetteville, AR 72701 USA (e-mail: soojo@uark.edu).

Guo-En Chang is with Department of Mechanical Engineering, National Chung Cheng University, Chia-Yi County 62102, Taiwan (e-mail: guoen1981@gmail.com).

Baohua Li is with Arktonics, LLC, 1339 South Pinnacle Drive, Fayetteville, AR 72701 USA (e-mail: baohua.li@arktonics.com).

Wei Du and Shui-Qing Yu are with Department of Electrical Engineering and Computer Science, University of Arkansas, Fayetteville, AR 72701 USA and the Institute for Nanoscience and Engineering, University of Arkansas, Fayetteville, AR 72701 USA (e-mails: weidu@uark.edu, syu@uark.edu).

Color versions of one or more of the figures in this article are available online at http://ieeexplore.ieee.org

passive devices to distribute the light on the chip, and it could allow for monolithically co-integrated photonic devices with electronics on the same chip, making such technology extremely attractive for broadening the scope of applications such as data processing, communication, sensing, etc [4–6]. Among many building blocks that make up a complete set of components for Si photonics, the most challenging task has been the efficient light sources such as lasers due to the indirect bandgap nature of Si. For the past few decades, while group III-V lasers have been adopted to be heterogeneously integrated on Si substrate via bonding or direct growth approaches [7, 8], Si photonics community keeps seeking for a monolithic integration technology. Recently, an emerging group-IV material, GeSn has been experimentally identified with direct bandgap as theoretically predicted [9], and therefore it holds great promise compared with group III-V materials to address the challenges of integration by offering monolithically integrated lasers on Si and CMOS compatible process. GeSn technology has attracted increasing attention rapidly not only because of the abovementioned merit but also due to its unique optical properties: the operating wavelength could cover broad short-wave and mid-wave infrared ranges [10]; by tuning Si and Sn compositions, the material lattice constant and bandgap energy can be engineered independently [11].

Using direct bandgap GeSn alloys, the optically pumped Fabry-Perot (F-P) cavity lasers were demonstrated in 2015 [12], with Sn composition of ~13% in active region, maximum lasing temperature of 90 K, and threshold of 325 kW/cm² at 20 K. Thanks to the enormous efforts on material growth that are boosting the laser development towards higher Sn incorporation to increase the bandgap directness [13, 14], growth of SiGeSn cladding layer to enhance the carrier confinement via double heterostructure (DHS) [15], and better interface control to reduce the defect density, the F-P cavity optically pumped GeSn lasers have achieved near room temperature operation (270 K) [16], lasing wavelength of 3 μm with 22.3% Sn incorporation [17], and continuously decreased lasing threshold. The distinct trend of improved device performance validates the development strategy and indicates the great potential of this gain material. In addition, the parallel work such as applying advanced optical cavities including microdisk to facilitate optical confinement and stressor to add tensile strain has been explored [18], showing reasonably improved device performance, as continuous-wave (CW) operation at 70 K and room temperature lasing.

The deep understanding of intrinsic material properties obtained from optical pumping study led to a breakthrough of the first electrically injected F-P cavity lasers reported in 2020 [19], with a maximum lasing temperature of 100 K, peak power of 2.7 mW/facet at 10 K, and thresholds of 0.6 and 1.4 kA/cm$^2$ at 10 and 77 K, respectively. Another demonstration of an electrically injected laser utilizes a microring cavity to promote strain relaxation [20]. The maximum lasing temperature of 90 K and threshold of 25 kA/cm$^2$ at 5 K were reported. Following the F-P cavity work, the loss mechanisms and device structure were comprehensively investigated [21], through which a clear pathway to substantially improve the carrier and optical confinement in the active region was provided. By implementing the optimized design, the higher lasing temperature of 135 K and reduced threshold of 0.8 kA/cm$^2$ at 77 K were obtained [22, 23]. While refs. [22, 23] only reports the preliminary results, in this work, the systemic study and detailed analysis are presented with more devices. The maximum lasing temperature was observed at 140 K, and the 77 K threshold was further reduced to 0.76 kA/cm$^2$. The improved device performance is attributed to: i) a thick cap layer to reduce the optical loss; ii) an appropriate SiGeSn barrier layer to enhance the carrier confinement; and iii) relatively narrow waveguide width to increase the optical confinement factor.

## II. EXPERIMENTAL METHODS

### A. Material Growth

The sample was grown using an industrial standard ASM Epsilon® 2000 Plus reduced pressure chemical vapor deposition (RPCVD) reactor with commercially available SiH$_4$, GeH$_4$, and SnCl$_4$ precursors. First, a 700-nm-thick phosphorous-doped Ge buffer layer was grown on an n-type Si(001) substrate by a two-step growth method [24]. Then a 600-nm-thick phosphorous-doped Ge$_{1-x}$Sn$_x$ buffer layer with the gradient Sn composition ranging from 8% to 12% was grown to reduce the compressive strain as well as to ensure a low density of threading dislocations propagating along the growth direction. This buffer also serves as an n-type contact layer. Note that the formation of Sn gradients is a spontaneous character due to enhanced Sn incorporation with strain relaxation [25], [26]. Subsequently, a 600-nm-thick undoped Ge$_{1-x}$Sn$_x$ active layer with 13% Sn composition and a 700-nm-thick boron-doped Si$_y$Ge$_{1-x-y}$Sn$_x$ cap layer with 8% Sn and 4% Si compositions were grown coherently strained to the Ge$_{1-x}$Sn$_x$ buffer. The sample structure is shown in Fig. 1(a).

### B. Structural Characterization

X-ray diffraction (XRD) measurements were carried out using a Panalytical X'Pert Pro MRD diffractometer equipped with a 1.8 kW Cu Kα1 X-ray tube (λ = 1.540598 Å), a standard four-bounce Ge(220) monochromator, and a Pixel detector. The cross-section transmission electron microscopy

(TEM) images were obtained using a FEI Tecnai microscope operated at an accelerating voltage of up to 200 kV. The optical properties were investigated through temperature-dependent photoluminescence (PL) performed using a standard off-axis configuration with a lock-in amplifier. A 1064 nm nanosecond pulsed laser was used as the excitation source. The PL emission was collected by a spectrometer and then sent to an InSb detector with a cutoff wavelength at 5.0 μm.

### C. Laser Diode Fabrication and Characterization

The sample was fabricated into a ridge-waveguide laser with ridge widths of 80 μm, 40 μm, and 20 μm, using standard photolithography and wet etching procedures. The etching depth of 1.5 μm was chosen to expose the n-type GeSn buffer layer for subsequent metal contacts. After etching, the samples were lapped down to 120 μm thick and then cleaved to form the Fabry-Perot cavity with a 1 mm length. To facilitate low-temperature measurements, the device was wire-bonded to a Si chip carrier.

The temperature-dependent current-voltage (IV) measurements were conducted utilizing a direct current (DC) source measurement unit (Keysight B2911). For light output versus injection current (LI) and spectra measurements, the laser device was driven by a pulsed high-compliance voltage source with a repetition rate of 1 kHz and a pulse width of 700 ns. The current was obtained using a calibrated magnetically coupled current meter. Electroluminescence (EL) emission was collected and analyzed using a monochromator with 10 nm resolution and a liquid-nitrogen-cooled InSb detector with a response range of 1-5.5 μm.

## III. RESULTS AND DISCUSSIONS

### A. Material Characterization

XRD, TEM, and SIMS measurements were used to examine and cross-check Si and Sn compositions, layer thickness, and strain of each layer. Figure 1(b) shows the XRD reciprocal space map (RSM) measured in the vicinity of ($\overline{2}\overline{2}4$) reflection of Ge. The peaks related to the Ge buffer, SiGeSn, and GeSn layers can be seen on the RSM from top to bottom in the order of increasing Sn composition. The close location of the peaks to the relaxation line (R = 100%) reveals a high degree of strain relaxation in each layer. In particular, the Ge buffer peak is on the right side of the line, which corresponds to 0.15% tensile strain originating from the different thermal expansion coefficients between Ge and Si [27], [28]. The location of the GeSn peaks on the left side of the relaxation line reveals a small compressive strain of less than -0.09% in the GeSn region, while the opposite location of the SiGeSn peaks corresponds to 0.07% tensile strain in the SiGeSn region resulting from the coherent growth on the underneath GeSn with a larger lattice constant. The RSM reveals that both the SiGeSn and GeSn peaks have a slight split, which is due to the different elemental composition in the buffer, active, and cap layers. In particular, according to the positions of the GeSn peaks, the Sn concentration is 12.9% in the active layer and gradually increases from 8 to 12% in the buffer layer along the growth direction. The TEM image shown in Fig. 1(c) reveals a high density of threading dislocations (TDs) in the GeSn buffer within a ~270 nm region from the GeSn/Ge interface and also confirms the subsequent growth of the low-defect GeSn active layer. The defect region is formed due to the gradual relaxation of mismatch strain during the GeSn layer growth directly on Ge [25].

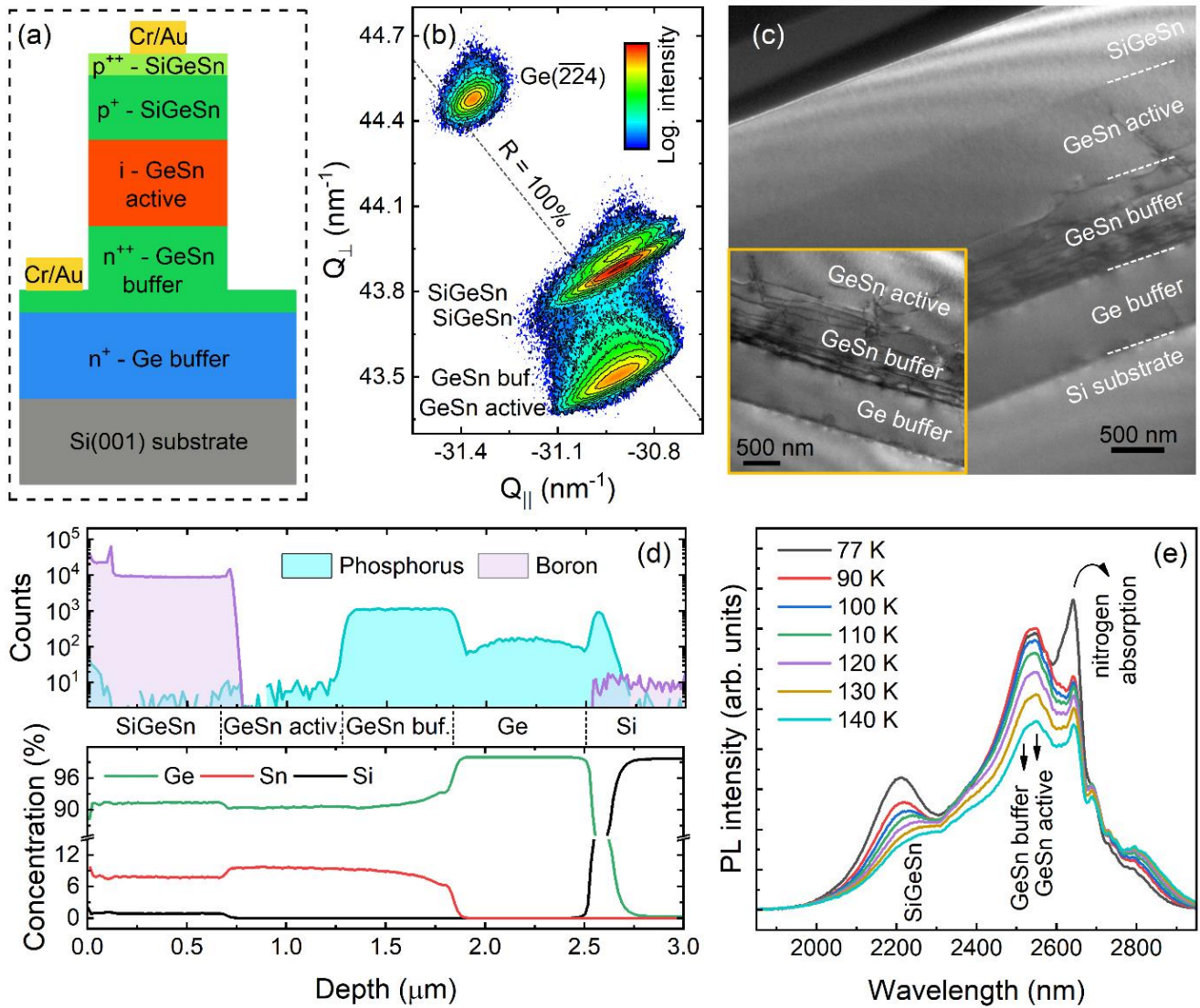

**Fig. 1.** (a) Schematic sample structure studied in this work. (b) X-ray diffraction ($\overline{2}\overline{2}4$) RSM showing the reciprocal lattice points of Ge, SiGeSn, and GeSn and a high degree of strain relaxation in each layer. (c) Cross-sectional TEM showing the high-quality GeSn active layer. (d) The SIMS profiles of doping concentration (top panel) and elemental composition (bottom panel). (e) Temperature-dependent PL excited under 1064 nm pumping laser.

The SIMS profiles were measured to examine the PIN structure as well as to cross-check the elemental composition of each layer. Figure 1(d) shows that the Sn concentration in the SiGeSn cap layer is 7.8%, while the Si concentration varies from 0.9% to 1.1% near the surface, which agrees with the peak splitting on the RSM. From the SIMS profiles, it also can be seen that the Sn concentration in the GeSn active and buffer layer is lower by about 1% as compared to XRD RSM data. Furthermore, the profiles of doping concentration confirm the PIN structure of the device, which consists of a boron-doped SiGeSn layer followed by the intrinsic GeSn active layer and phosphorous-doped GeSn and Ge buffer layers.

Figure 1(e) shows the temperature-dependent (TD) PL spectra. According to bandgap energy determined by Si and Sn compositions, the shorter wavelength peak at 2214 nm is assigned to emission from the SiGeSn barrier; the longer wavelength peak at 2538 nm corresponds to emissions from the GeSn active region and the top of the GeSn buffer where the Sn composition is 12%. Since GeSn alloy with higher Sn composition has narrower bandgap energy and the photo-generated carriers tend to flow towards the narrow bandgap area, the recombination mainly occurs at the higher Sn region.

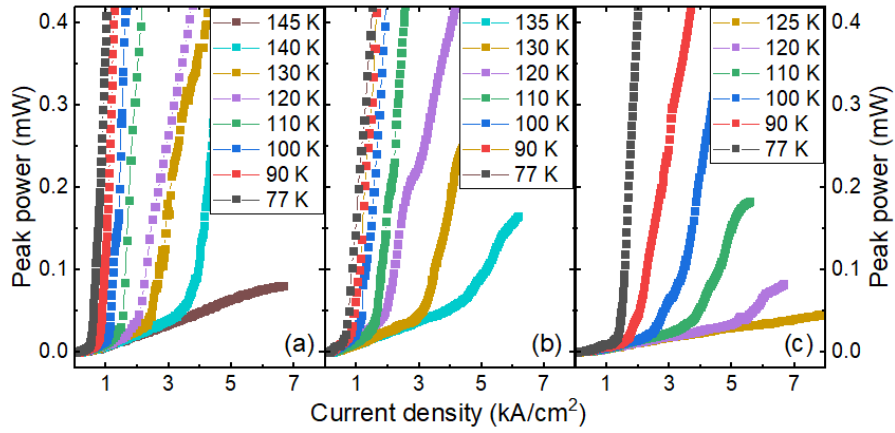

**Fig. 2.** Temperature-dependent LI curves of (a) 80 μm, (b) 40 μm, and (c) 20 μm wide ridge devices.

The peak at 2642 nm is due to nitrogen absorption [29]. The increased PL intensity at lower temperatures was observed [12], and the peak shifts towards shorter wavelengths as expected. Note that for highly doped GeSn buffer and SiGeSn cap layers, the PL peaks are at longer wavelengths compared to their intrinsic counterparts, agreeing with the bandgap narrowing effect [30].

### B. Laser Device Characterization

Figure 2 shows the laser output power-current (LI) curves for 80 μm, 40 μm, and 20 μm wide ridge devices measured from 77 K to the maximum operating temperature. In Fig. 2(a), a clear kink on the LI curve of the 80 μm wide ridge device was observed up to 140 K, and the threshold is as low as 0.756 kA/cm² at 77 K. The maximum lasing temperature is higher than that in the previous study of a similar size device while the threshold current density is much lower (1.4 kA/cm² in the previous study). This is due to that: 1) utilizing a 700-nm-thick SiGeSn cap dramatically reduces the optical loss arising from metal contact. Although the heavily doped SiGeSn introduces free carrier absorption (FCA) loss, the overall optical and FCA loss would reach a minimum value at a certain thickness, which is ~700 nm based on numerical calculation [31]. The improved device performance in this work agrees well with previous theoretical study. 2) adopting a SiGeSn barrier effectively enhances the carrier confinement, which offers a higher barrier height in the conduction band with respect to the GeSn active layer. At the valence band, the hole leakage issue due to type-II band alignment in the light hole band between the GeSn active and SiGeSn cap as a result of tensile strain is addressed by injecting holes from the top SiGeSn cap layer. The LI curves reveal lower temperatures of laser operation for narrower ridge devices, as shown in Fig. 2(b) and (c). The maximal lasing temperatures decrease to 135 K and 120 K for 40 μm and 20 μm devices, respectively. This can be explained by that although a narrower device features better optical confinement compared to wider ridge devices, it suffers from worse heat dissipation due to the smaller surface area. In addition, the surface roughness leading to scattering loss would be more pronounced with a narrow ridge device.

At 77 K, the threshold current densities were extracted as 0.858 kA/cm² and 1.24 kA/cm², respectively. The performance of all three devices is superior to previous ones attributed to optimized structural design. The lasing characteristics of all devices are listed in Table I.

| Device ridge width | Max lasing temp (K) | Peak power (mW/facet) | Lasing wavelength (nm) | | Threshold current (A) | Threshold current density (kA/cm²) | |
|---|---|---|---|---|---|---|---|
| | | | At 77 K | At max lasing temp | | At 77 K | At max lasing temp |
| 80 μm | 140 | 2.20 | 2654 | 2722 | 0.614 | 0.756 | 4.10 |
| 40 μm | 135 | 1.15 | 2604 | 2658 | 0.336 | 0.858 | 4.54 |
| 20 μm | 120 | 0.85 | 2652 | 2678 | 0.295 | 1.512 | 6.06 |

Figure 3 shows the EL spectra below (0.9 ×), at (1.0 ×), and above (1.1 ×) lasing threshold (Jth) measured at 77 K and at the corresponding maximum lasing temperature for each device. At 0.9 × Jth excitation level, only a broad emission peak corresponding to spontaneous emission was obtained, which can be identified to be from GeSn buffer and active region according to PL study. At the threshold, an emerging peak at ~2.7 μm was clearly observed, suggesting the onset of lasing from the GeSn active region. As the injection level increases from 1.0 × to 1.1 × Jth, this peak grows rapidly and sharply. The significantly reduced peak line width unambiguously indicates the stimulated emission, showing clear evidence of laser action. For all three devices, with increased temperature, the lasing peak shifts towards longer wavelengths as expected.

The emission power was measured at 77 K and plotted in Fig. 4. The maximum peak power of 2.2 mW/facet was obtained from the 80 μm device at the injection current of 2 mA (3.02 kA/cm²), much higher than that for the previously reported device with similar size (1.2 mW/facet at 6.0 kA/cm²). The current was not further increased to prevent the device failure. The external quantum efficiency (EQE) was estimated as less than 1%. The relatively low EQE can be interpreted as follows: i) SiGeSn cap and GeSn buffer serve as top and bottom barriers, however the barrier heights are still insufficient, which creates carrier leakage channel and therefore considerably reduces the carrier injection efficiency; 2) using a thick SiGeSn cap could reduce the optical loss. On the other hand, the gradually relaxed material would bring dislocations creating defect energy levels in the bandgap, leading to additional absorption loss; and 3) the F-P cavities were fabricated using simply wet etching, and the facet was formed by manual cleaving, and therefore the scattering loss due to surface roughness further degrades the efficiency.

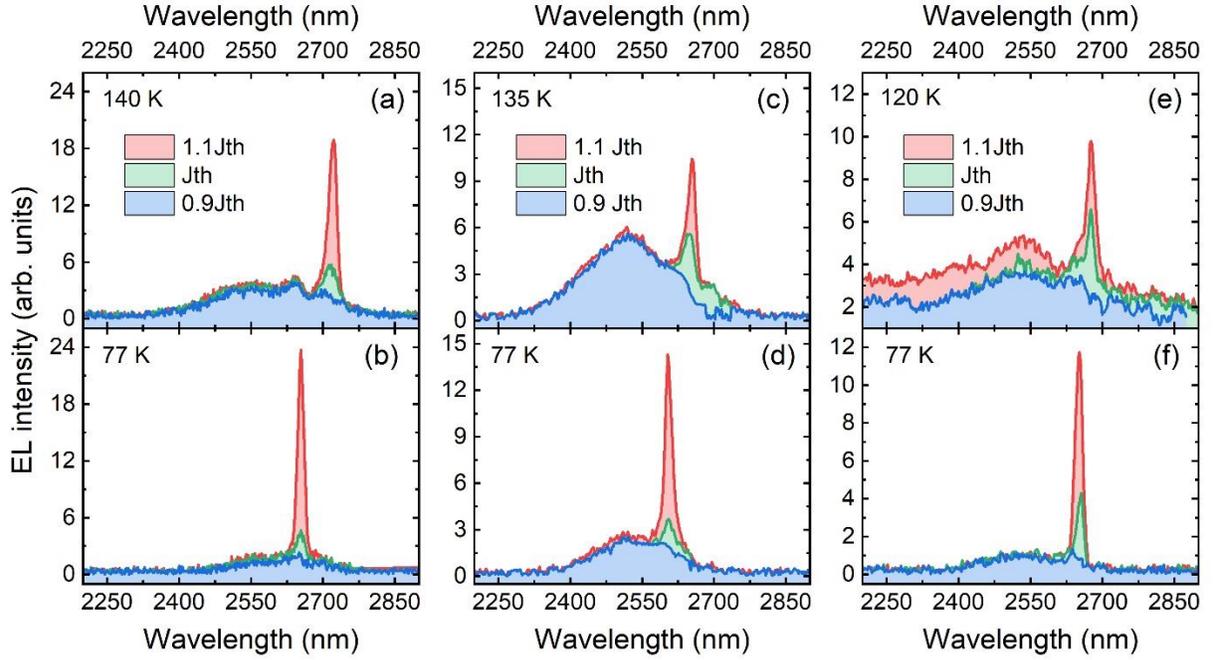

**Fig. 3.** Emission spectra at different injection levels at (a) 140 K and (b) 77 K for 80 μm width device; (c) and (d) for 40 μm width device; (e) and (f) for 20 μm width device.

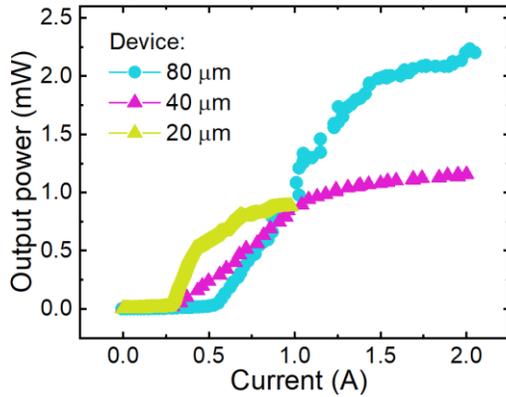

**Fig. 4.** Emission power measured at 77 K for 80 μm, 40 μm, and 20 μm wide ridge devices.

Improvement of laser performance can be carried out by addressing the abovementioned issues: growing a lattice-matched SiGeSn cap (with respect to GeSn active layer) with a wider bandgap to reduce the density of dislocations and enhance carrier confinement; inserting the same SiGeSn between GeSn active and buffer to mitigate the carrier leakage; optimizing fabrication procedure to reduce the scattering loss.

## IV. CONCLUSION

We have demonstrated electrically injected GeSn/SiGeSn lasers emitting at 2.7 μm. By using a thick SiGeSn cap layer, which also serves as a barrier, the overall optical loss was reduced, and optical field and carrier confinement were enhanced, leading to dramatically improved device performance. The maximum lasing temperature of 140 K and low threshold of 0.756 kA/cm² at 77 K were achieved. The measured peak power was 2.2 mW/facet at 77 K. Further improvement of laser performance can be achieved by optimizing the cap layer design and fabrication procedure.

### ACKNOWLEDGMENT

The authors are grateful for the support from Sylvester Amoah at University of Arkansas, Fayetteville on laser mask design and measurement.

### REFERENCES

[1] K. P. Homewood and M. A. Lourenço, "The rise of the GeSn laser," *Nature Photonics*, vol. 9, no. 2. Nature Publishing Group, pp. 78–79, 17-Feb-2015.

[2] J. J. Ackert *et al.*, "High-speed detection at two micrometres with monolithic silicon photodiodes," *Nat. Photonics*, vol. 9, no. 6, pp. 393–396, 2015.

[3] C. Sun *et al.*, "Single-chip microprocessor that communicates directly using light," *Nature*, vol. 528, no. 7583, pp. 534–538, 2015.

[4] G. Roelkens *et al.*, "III-V-on-Silicon Photonic Devices for Optical Communication and Sensing," *Photonics*, vol. 2, no. 3, pp. 969–1004, 2015.

[5] M. Tang *et al.*, "Integration of III-V lasers on Si for Si photonics," *Prog. Quantum Electron.*, vol. 66, pp. 1–18, 2019.

[6] Z. Wang *et al.*, "Novel Light Source Integration Approaches for Silicon Photonics," *Laser \& Photonics Rev.*, vol. 11, no. 4, p. 1700063, 2017.

[7] G. Roelkens *et al.*, "III-V/silicon photonics for on-chip and intra-chip optical interconnects," *Laser \& Photonics Rev.*, vol. 4, no. 6, pp. 751–779, 2010.

[8] M. Takenaka *et al.*, "Heterogeneous CMOS Photonics Based on SiGe/Ge and III–V Semiconductors Integrated on Si Platform," *IEEE J. Sel. Top. Quantum Electron.*, vol. 23, no. 3, pp. 64–76, 2017.

[9] S. A. Ghetmiri *et al.*, "Direct-bandgap GeSn grown on silicon with 2230 nm photoluminescence," *Appl. Phys. Lett.*, vol. 105, no. 15, p. 151109, Oct. 2014.

[10] H. Tran *et al.*, "Si-Based GeSn Photodetectors toward Mid-Infrared Imaging Applications," *ACS Photonics*, vol. 6, no. 11, pp. 2807–2815, Nov. 2019.

[11] O. Olorunsola *et al.*, "SiGeSn Quantum Well for Photonics Integrated Circuits on Si Photonics Platform: A Review," *J. Phys. D. Appl. Phys.*, 2022.

[12] S. Wirths *et al.*, "Lasing in direct-bandgap GeSn alloy grown on Si," *Nat. Photonics*, vol. 9, no. 2, pp. 88–92, Feb. 2015.

[13] N. Von Den Driesch *et al.*, "Direct Bandgap Group IV Epitaxy on Si for Laser Applications," *Chem. Mater.*, vol. 27, no. 13, pp. 4693–4702, 2015.

[14] J. Aubin *et al.*, "Growth and structural properties of step-graded, high Sn content GeSn layers on Ge," *Semicond. Sci. Technol.*, vol. 32, no. 9, p. 094006, Sep. 2017.

[15] N. von den Driesch *et al.*, "Advanced GeSn/SiGeSn Group IV Heterostructure Lasers," *Adv. Sci.*, vol. 5, no. 6, p. 1700955, 2018.

[16] Y. Zhou *et al.*, "Optically Pumped GeSn Lasers Operating at 270 K with Broad Waveguide Structures on Si," *ACS Photonics*, vol. 6, no. 6, pp. 1434–1441, Jun. 2019.

[17] W. Dou *et al.*, "Optically pumped lasing at 3 μm from compositionally graded GeSn with tin up to 22.3%," *Opt. Lett.*, vol. 43, no. 19, pp. 4558–4561, Oct. 2018.

[18] A. Elbaz *et al.*, "Ultra-low-threshold continuous-wave and pulsed lasing in tensile-strained GeSn alloys," *Nat. Photonics*, vol. 14, no. 6, pp. 375–382, 2020.

[19] Y. Zhou *et al.*, "Electrically injected GeSn lasers on Si operating up to 100 K," *Optica*, vol. 7, no. 8, pp. 924–928, Aug. 2020.

[20] B. Marzban *et al.*, "Strain Engineered Electrically Pumped SiGeSn Microring Lasers on Si," *ACS Photonics*, vol. 10, no. 1, pp. 217–224, Jan. 2023.

[21] Q. M. Thai *et al.*, "GeSn optical gain and lasing characteristics modelling," *Phys. Rev. B*, vol. 102, no. 15, p. 155203, Oct. 2020.

[22] S. Ojo, Y. Zhou, S. Acharya, N. Saunders, S. Amoah, Y.-T. Jheng, H. Tran, W. Du, G.-E. Chang, B. Li, S.-Q. Yu, "Silicon-based electrically injected GeSn lasers," CLEO, accepted, May 5-10, 2024, Charlotte, North Carolina.

[23] S. Acharya, S. Ojo, Y.Zhou, S. Amoah, W.Du, B. Li, S. Yu, "Electrically injected GeSn laser on Si substrate operating up to 130 K," MIOMD - XVI Mid-IR Optoelectronics: Materials and Devices Sessions, MIOMD-TuM2, August 6-10, 2023, Norman, OK.

[24] J. Margetis *et al.*, "Growth and Characterization of Epitaxial Ge₁₋ₓSnₓ Alloys and Heterostructures Using a Commercial CVD System," *ECS Trans.*, vol. 64, no. 6, pp. 711–720, Aug. 2014.

[25] W. Dou *et al.*, "Investigation of GeSn Strain Relaxation and Spontaneous Composition Gradient for Low-Defect and High-Sn Alloy Growth," *Sci. Rep.*, vol. 8, no. 1, p. 5640, Dec. 2018.

[26] S. Assali, J. Nicolas, and O. Moutanabbir, "Enhanced Sn incorporation in GeSn epitaxial semiconductors via strain relaxation," *J. Appl. Phys.*, vol. 125, no. 2, p. 025304, Jan. 2019.

[27] J. M. Hartmann *et al.*, "Reduced pressure-chemical vapor deposition of Ge thick layers on Si(001) for 1.3-1.55-μm photodetection," *J. Appl. Phys.*, vol. 95, no. 10, pp. 5905–5913, May 2004.

[28] D. D. Cannon *et al.*, "Tensile strained epitaxial Ge films on Si(100) substrates with potential application in L-band telecommunications," *Appl. Phys. Lett.*, vol. 84, no. 6, pp. 906–908, Feb. 2004.

[29] O. Olorunsola *et al.*, "Investigation of SiGeSn/GeSn/SiGeSn single quantum well with enhanced well emission," *Nanotechnology*, vol. 33, no. 8, p. 85201, Nov. 2021.

[30] R. Camacho-Aguilera, Z. Han, Y. Cai, L. C. Kimerling, and J. Michel, "Direct band gap narrowing in highly doped Ge," *Appl. Phys. Lett.*, vol. 102, no. 15, p. 152106, 2013.

[31] Y. Zhou *et al.*, "Electrically injected GeSn lasers with peak wavelength up to 2.7 μm," *Photon. Res.*, vol. 10, no. 1, pp. 222–229, Jan. 2022.